\documentclass[twocolumn,prl,showpacs,amsmath,amssymb,nofootinbib,superscriptaddress,preprintnumbers]{revtex4}

\usepackage{dcolumn}
\usepackage{amsmath}
\usepackage{graphicx}
\usepackage{rotating}
\usepackage{amsfonts}
\usepackage{amssymb}
\usepackage{subfigure}
\usepackage{slashed}

\usepackage{graphicx}

\newcommand{\mix}{\chi}
\newcommand{\mixe}{\chi_\mathrm{eff}}
\newcommand{\OP}{\omega_\mathrm{P}}

\newcommand{\muu}{ m_{\gamma^\prime}}
\newcommand{\R}{\mathrm{res}}
\newcommand{\G}{\Gamma_\mathrm{C}}
\newcommand{\gp}{\gamma^{\prime}}
\newcommand{\rr}{\mathrm{R}}
\newcommand{\rrr}{\mathrm{r}}

\begin{document}

\preprint{IPPP/08/28}
\preprint{DCPT/08/56}
\preprint{DESY 08-045}

\title {Signatures of a hidden cosmic microwave background}
\author{Joerg Jaeckel}
\affiliation{Institute for Particle Physics and Phenomenology, Durham University, Durham DH1 3LE, UK}
\author{Javier Redondo}
\affiliation{Deutsches Elektronen-Synchrotron DESY, Notkestra\ss e 85, D-22607 Hamburg, Germany}
\author{Andreas Ringwald}
\affiliation{Deutsches Elektronen-Synchrotron DESY, Notkestra\ss e 85, D-22607 Hamburg, Germany}

\begin{abstract}
If there is a hidden photon  -- i.e. a light abelian gauge boson $\gp$ in the hidden sector
-- its kinetic mixing with the standard photon
can produce a hidden cosmic microwave background (hCMB).
For meV masses, resonant oscillations $\gamma\leftrightarrow \gp$ happen after nucleosynthesis (BBN)
but before CMB decoupling,
increasing the effective number of neutrinos ($N_\nu^\mathrm{eff}$) but also the baryon to photon ratio
at decoupling.
The current agreement  between BBN and CMB data provides new constraints on the kinetic mixing.
However, if one includes Lyman-$\alpha$ data, $N_\nu^\mathrm{eff}>3$ is preferred.
It is tempting to interpret this effect in terms of the hCMB.
Interestingly, the required parameters will be tested in the near future by laboratory experiments.
\end{abstract}


\maketitle

Most embeddings of the standard model into a more unified
theory, in particular the ones based on supergravity or
superstrings, predict the existence of a hidden sector whose inhabitants
have only very weak interactions with the standard model.
The gauge interactions in the hidden sector generically involve U(1) factors.
Usually, it is assumed that the corresponding gauge bosons are very
heavy, in order to avoid observational constraints.
However, in realistic string compactifications,
one of these hidden photons may indeed be light, with a mass in the
sub-eV range, arising from a Higgs or St\"uckelberg mechanism.
In this case, the dominant interaction with the visible sector photon will be through
gauge kinetic mixing~\cite{Holdom:1985ag}, i.e. the
system can be parametrized by the low-energy effective Lagrangian,
\begin{eqnarray}
\mathcal{L} &=&
-\frac{1}{4}F_{\mu \nu} F^{\mu \nu}
 - \frac{1}{4}B_{\mu \nu} B^{\mu \nu}
+ \frac{\sin\chi}{2} B_{\mu \nu} F^{\mu \nu}
\\\nonumber
&&\quad\quad\quad\quad\quad\quad\quad\quad\quad\quad\quad\quad
+ \frac{\cos^2\mix}{2} \muu^2 B_{\mu} B^{\mu},
\label{LagKM}
\end{eqnarray}
where $F_{\mu\nu}$ and $B_{\mu\nu}$ are the photon ($A^\nu$) and hidden photon ($B^\nu$) field strengths.
The dimensionless mixing parameter $\sin\chi$ can be generated at an arbitrarily high
energy scale and does not suffer from any kind of mass suppression from the messenger particles
communicating between the visible and the hidden sector. This makes it an extremely powerful probe
of high scale physics. Typical predicted values for $\chi$ in realistic string compactifications
range between $10^{-16}$ and $10^{-2}$~\cite{Dienes:1996zr}.

The most prominent implication of the kinetic mixing term is that photons are no longer massless
propagation modes.
Similar to neutrino mixing the propagation and the interaction eigenstates are misaligned.
The kinetic mixing term can be removed by changing the basis $\{A,B\}\rightarrow\{A_{_R},S\}$,
where $A_{_R}=\cos\mix A$, $S=B-\sin\mix A$.
Since $A$ and $A_{_R}$ differ only by a typically unobservable charge renormalization we
will drop the $R$ subscript from now on.
In the $\{A,S\}$ basis the kinetic term is diagonal but kinetic mixing provides an off-diagonal term
in the mass-squared matrix,
\begin{equation}
\left(
\begin{array}{cc}
\muu^2 \sin\mix^2 & \muu^2  \sin\mix\cos\mix     \\
\muu^2\sin\mix\cos\mix   & \muu^2 \cos^2\mix
\end{array}
\right) \ \ .
\end{equation}
As a result one expects vacuum photon-sterile oscillations~\cite{Okun:1982xi} as in the case of $\nu_e-\nu_\mu$.

In this letter, we examine the implications of this simple scenario for late cosmology.
We focus on the meV mass range where a thermal population of hidden photons can be created through
resonant oscillations after BBN but before CMB decoupling.
This `hidden CMB' (hCMB) will contribute to the effective number of additional neutrinos at decoupling and,
since some photons will disappear, will increase the baryon to photon ratio with respect to the BBN value.
Moreover, we will argue that possible distortions of the thermal nature of the CMB spectrum are very small
in the parameter range allowed by other constraints.\\

\noindent\emph{Effects of a post BBN hCMB production.-}
Let us assume that at the time of BBN there is no hidden photon thermal bath present.
For the case where kinetic mixing is the only interaction with standard model particles,
this will be justified later.
Additional light particles charged under the hidden U(1)$_\mathrm{h}$ would appear as
minicharged particles (MCPs) which could mediate the formation of a hCMB before BBN.
Given the existent constraints on the number of additional neutrinos at BBN and
on MCPs~\cite{Davidson:2000hf} we will not discuss this possibility here and focus instead on the
case when there is only a hidden photon.

The oscillation of photons into hidden-sector photons decreases the photon number and energy density
($n_\gamma, \rho_\gamma$) leaving the total energy unchanged.  The key parameter will be the fraction
of $\rho_\gamma$ that is converted into hidden photons, $x\equiv\rho_{\gamma'}/\rho_\gamma$.

We will see that inelastic processes are effective after hCMB decoupling. Therefore the remaining photons
will regain a black body distribution, albeit, due to the energy loss, at a lower temperature,
\begin{equation}
T_{\rm after}=(1-x)^{\frac{1}{4}}T_{\rm before} \ .
\end{equation}

Since neutrinos remain unchanged during the hCMB formation, the ratio of neutrino and photon temperatures
will also increase.
The invisible energy density (in radiation) at decoupling can be estimated using CMB anisotropies and is
often quoted
as the effective number of `standard' neutrinos, by normalizing it with $(7/8)(4/11)^{4/3}\rho_\gamma$.
In our case,
\begin{equation}
N_\nu^\mathrm{eff}\equiv
\frac{\rho^\mathrm{total}-\rho_\gamma}{\frac{7}{8}\left(\frac{4}{11}\right)^\frac{4}{3} \rho_\gamma}
= \frac{N_\nu^\mathrm{std}}{(1-x)^{}} + \frac{x}{1-x}\frac{8}{7}\left(\frac{11}{4}\right)^\frac{4}{3}
\end{equation}
is the sum of neutrino and hidden photon contributions.

Strong limits on $N_\nu^\mathrm{eff}$ at decoupling arise from global fits of CMB anisotropies
alone \cite{Ichikawa:2008pz}
or combined with large scale structure data \cite{Mangano:2006ur}. Inclusion of Lyman-$\alpha$ data
favors values $N_\nu^\mathrm{eff}>3$ \cite{Seljak:2006bg}, although this might well be due to
systematics \cite{Hamann:2007pi}. A recent analysis of WMAP5 plus other CMB anisotropy probes,
large scale structure (no Ly-$\alpha$) and supernovae data provides \cite{Simha:2008zj}
\begin{equation}
N^{\rm eff}_{\nu}=2.9^{+2.0}_{-1.4} \quad\,(95\%\,\,\mathrm{C.L.})\ \      ,  \label{Neff}
\end{equation}
which, using the standard, $N^{\rm std}_{\nu}=3.046$, turns \eqref{Neff} into
\begin{equation}
x\lesssim 0.20 \ .
\end{equation}
As a conservative example of the inclusion of Ly-$\alpha$ data we propose
$N^{\rm eff}_\nu  =3.8^{+2.0}_{-1.6}\,(95\%\,\,\mathrm{C.L.})$  \cite{Hamann:2007pi},
suggesting $x \simeq 0.1$. It is worth noting however that values as high as $N^{\rm eff}_\nu=5.3$
can be found also in the literature \cite{Seljak:2006bg}.

Since $n_\gamma$ is proportional to $T^3$, the baryon to photon ratio (which would otherwise remain constant)
is also modified according to
\begin{equation}
\eta^{\rm after}=(1-x)^{-\frac{3}{4}} \eta^{\rm before}\ .
\end{equation}
Indeed the value for $\eta$ inferred from the abundances of the light elements produced at BBN
and the one obtained by measuring temperature fluctuations in the CMB \cite{Simha:2008zj}  agree
within their error bars,
\begin{eqnarray}
&&\!\!\!\eta^{\rm BBN}=5.7^{+0.8}_{-0.9} \times 10^{-10}\quad \,\,\,\,(95\%\,\, \mathrm{C.L.}),  \label{etaBBN}
\\\
&&\!\!\!\eta^{\rm CMB}=6.14^{+0.3}_{-0.25} \times 10^{-10}\quad \,\,\,\,(95\%\,\, \mathrm{C.L.}),
\end{eqnarray}
which allows us to set the bound $x\lesssim 0.32$.\\


\noindent \emph{Photon oscillations in the early universe plasma.-}
The formalism to study the dynamics of a thermal bath of particles that undergo `flavor' oscillations
among the different species was developed some time ago \cite{Harris:1980zi,Stodolsky:1986dx}, for a
textbook treatment see \cite{Raffelt:1996wa}.

The state of an ensemble of $\gamma$ and $\gp$ is described by a 2-by-2 density matrix,
\begin{equation}
\rho =\frac{1}{2}(1+{ \mathbf P \cdot \boldsymbol\sigma }) \, ,
\end{equation}
where $\boldsymbol\sigma$ has the Pauli matrices as components and $ \mathbf{P} $ is a
`flavor polarization vector' carrying all the information of the ensemble. The modulus $|\mathbf{P}|$
gives the degree of coherence,
$\mathrm{P}_z=1$ $(-1)$ corresponding to a pure $\gamma$ ($\gp$) state, while $|\mathbf{P}|=0$ to a
completely incoherent state, which of course defines the state of `flavor equilibrium'.
The transverse components $\mathrm{P}_{x,y}$ contain the quantum correlations.

The time evolution of the ensemble is given by a precession of $\mathbf{P}$ (flavor oscillations)
and a shrinking of its transverse component (decoherence due to absorption and
scattering), according to  Stodolsky's formula \cite{Stodolsky:1986dx}
\begin{equation}
\dot{\mathbf{P}}=\mathbf{V}\times\mathbf{P}-D\mathbf{P}_\mathrm{T}\ .  \label{sto}
\end{equation}
Here $\mathbf{V}$ is the `flavor magnetic field' given by
\begin{equation}
\Delta\left(
\begin{array}{c} \sin2\mixe\\ 0\\ \cos2\mixe \end{array} \right) \equiv
\frac{\muu^2}{2\omega}\left( \begin{array}{c} \sin2\mix\\ 0\\ \cos2\mix \end{array} \right) -
\frac{\OP^2}{2\omega}\left( \begin{array}{c} 0\\ 0\\1\end{array} \right) \ ,
\end{equation}
where $\omega$ is the $\gamma,\gp$ energy and we have included the refraction properties of the medium
(basically an electron plasma)
as a photon `effective mass', given by the plasma frequency $\OP^2 \simeq 4 \pi \alpha n_e/m_e$
with $\alpha$ the fine-structure constant, and $m_e, n_e$ the electron
 mass and density.

The damping factor $D$ equals half the collision rate of photons \cite{Stodolsky:1986dx}, here dominated
by Thomson scattering, $\G \simeq 8\pi\alpha^2/(3m_e^2)n_e$.

Before dealing with the details of the calculation, we will try to gain some intuition about the
main points of the cosmology of the $\gamma-\gp$ system.
For temperatures below $0.04\ m_e\sim 20$ keV, electrons and positrons have annihilated leaving an
electron relic density which balances the charge of protons,
namely $n_e\simeq n_B=\eta\ n_\gamma = \eta\ 2 \zeta(3) T^3/\pi^2$. Using the BBN central
value \eqref{etaBBN}, we find
\begin{equation}
\OP^2 \simeq (0.16\ \mathrm{meV})^2 \left(\frac{T}{\mathrm{keV}}\right)^3 \ \ . \label{OPT}
\end{equation}
As the universe expands, the density decreases and so do the plasma mass and absorption rate.
The ratio $\rr\equiv \omega\G/\OP^2$ is, however, independent of $n_e$ and very small,
$2\alpha\omega/(3m_e) \sim 10^{-5}(\omega/\mathrm{keV})$.
Therefore the damping term $D$ in \eqref{sto} is typically smaller than the precession rate $\Delta$,
the only possible exception being the resonant case to be discussed later.
In this situation we can use the precession-averaged $\mathbf{P}_\mathrm{T}$ in \eqref{sto}, which then turns
into $\langle \dot{|\mathbf{P}|}/|\mathbf{P}|\rangle=-\cos^22\mixe\sin^22\mixe\G/2$ \cite{Raffelt:1996wa}.
Under these conditions a significant hCMB will form if the rate
\begin{equation}
\tilde\Gamma\equiv \frac{1}{2}\cos^22\mixe\sin^22\mixe \G      \label{weakD}
\end{equation}
exceeds the expansion rate of the universe, given by the Hubble parameter
$H=2.5\times 10^{-22}\ (T/\mathrm{keV})^2$ eV.

\begin{figure}
\begin{center}
\includegraphics[width=7.5cm]{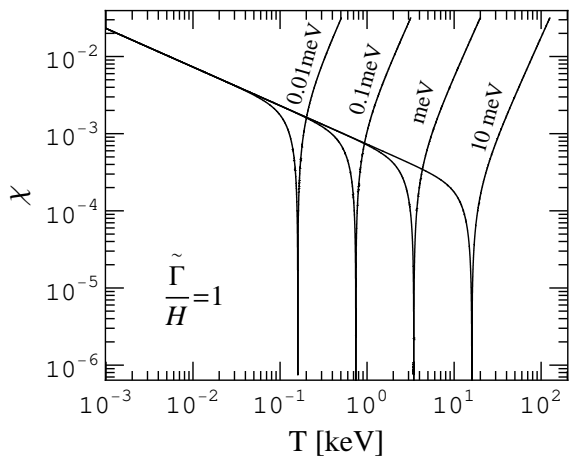}
\end{center}
\vspace{-5ex}
\caption{Isocontours of $\tilde\Gamma/H=1$ for different $\gp$ masses.\vspace{-3ex}}
\label{GoverH}
\end{figure}

Isocontours of $\tilde\Gamma/H=1$ have been plotted in Fig. \ref{GoverH}, showing the possible history
of the hCMB formation for
different values of $\muu$ and $\sin\mix$.
During its evolution, the universe moves on a horizontal line of fixed $\chi$ from high temperature (right)
to low temperatures (left).
We can clearly distinguish three different regimes  separated in time by a resonant peak occurring at $T^\R$,
the temperature at which $\OP^2$ has decreased enough to match exactly $\muu^2\cos2\mix$.

a) In an early stage, for $T\gg T^\R$, $\OP$ is so large that the hidden photon is comparatively massless;
photons are very
close to being both interaction and propagation eigenstates and $\gamma-\gp$ oscillations are strongly
suppressed by the effective mixing angle given by
\begin{equation}
\sin2\chi_\mathrm{eff}\simeq  \frac{\muu^2}{\omega_P^2}\sin2\chi \ll \sin2\chi \ \ .
\end{equation}
This is in contrast to the usual cosmology of exotic particles. Light hidden photons are completely
decoupled for
high enough temperatures (as long as their only relevant coupling is the kinetic mixing).
This justifies our earlier claim and allows us to set the initial
conditions for the $\gamma-\gp$ system to $\mathrm{P}_z=+1$, i.e. a pure photon bath without
hidden photons.
As the temperature approaches $T^\R$, $\sin2\mixe$ increases and can even be enhanced
compared to $\sin2\mix$.
Hidden photons will be effectively produced when crossing the $\tilde\Gamma/H=1$ line
in Fig. \ref{GoverH}.

b) In a small region around $T^\R$, the production can be effective even for very small $\mix$.
Unfortunately, the precession average is not justified because typically
$\Delta^\R \equiv\muu^2\sin2\mix/(2\omega)\ll D$.
Below we provide details on the calculation in this regime.

c) For $T\ll T^\R$, the averaging procedure is again justified.
In this regime, $\muu\ll \OP$ and we recover the well known vacuum case.
The evolution will again freeze out after crossing the $\tilde \Gamma/H=1$ line
in Fig. \ref{GoverH}.
\\

\begin{figure}
\begin{center}
\includegraphics[width=8.6 cm]{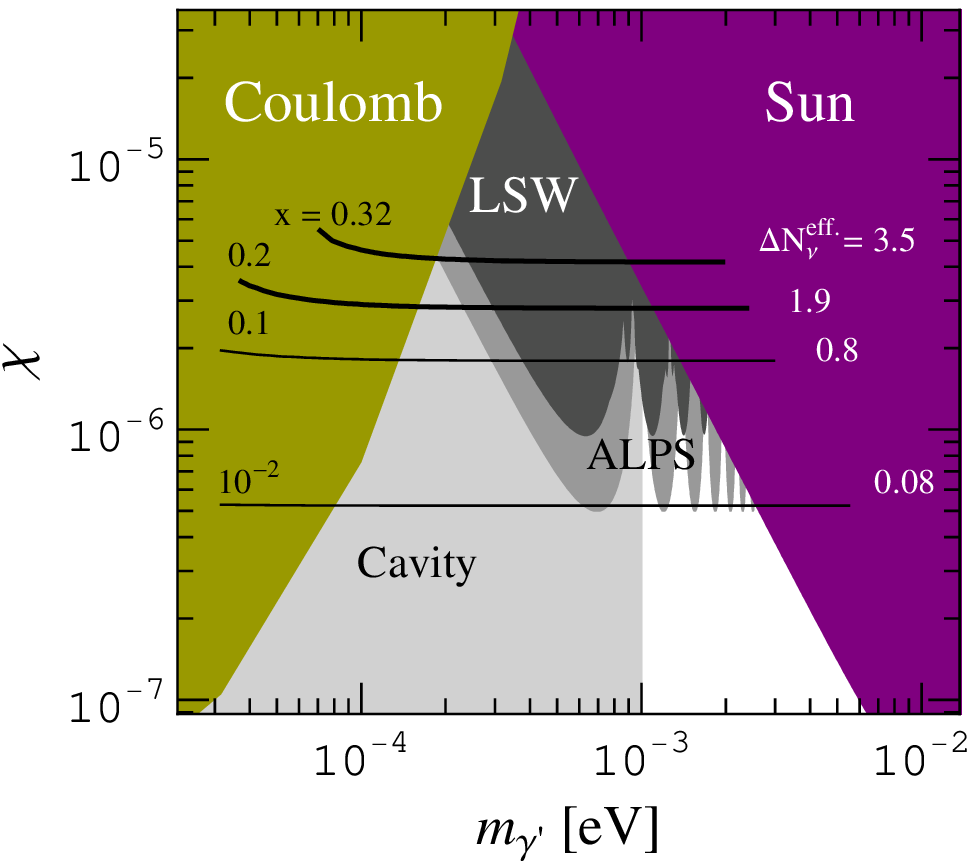}
\end{center}
\vspace{-5ex}
\caption[...]{
Isocontours of $x\equiv\rho_{\gamma'}/\rho_\gamma$ at the CMB epoch in the mass-mixing plane.
The region above $x=0.32$ is excluded by the agreement between the baryon to photon ratio inferred
from BBN and CMB data, while
 $x>0.2$ is excluded by the upper limits on the cosmic radiation density at CMB, usually expressed in
terms of the effective number of neutrinos $N_\nu^\mathrm{eff}=3.046+\Delta N_\nu^\mathrm{eff}$.
Also shown are the exclusion bounds from Coulomb law tests \cite{Bartlett:1988yy},
CAST (Sun) \cite{Redondo:2008aa}
and light-shining-through-walls (LSW) experiments \cite{Ahlers:2007qf}.
The near future prospects of the ALPS
experiment are also shown \cite{ALPS}. Sensitivity in the region labeled `Cavity' can be obtained by an
experiment using microwave cavities \cite{Jaeckel:2007ch}. The remaining region at higher masses can be
probed by solar hidden photon searches \cite{Gninenko}.\vspace{-3ex}
\label{Prob}}
\end{figure}

\noindent\emph{Resonant production.-}
>From Fig. \ref{GoverH} it is clear that for $\chi\lesssim 6\times
10^{-6}$ (larger values are excluded by laboratory experiments, see
Fig. \ref{Prob}) production is effective only in the resonant
regime. Therefore, we need an approximation that is valid in the
vicinity of the resonance. Moreover, since the resonance happens
only for a short period of time, the simple criterion $\tilde
\Gamma/H>1$ is not sufficient to ensure that a sizable hCMB is
produced. Accordingly we have to calculate the integrated
production.

In the vicinity of the resonance, the oscillation frequency is minimal
($\Delta^\R\equiv\muu^2\sin2\mix/(2\omega)$) and indeed typically much smaller than the
damping factor $\G$.
The evolution of $\mathbf{P}$ in a general `strong damping' regime was studied in \cite{Stodolsky:1986dx}
where it was found that the flavor relaxation rate is given by
\begin{equation}
\overline{\Gamma}=V_\mathrm{T}^2\frac{D}{D^2+V_\mathrm{z}} =
\frac{1}{2} \frac{\sin^22\mix}{\rrr^2+(\cos2\mix-y)^2} \G \ ,
\label{strongD}
\end{equation}
with $y\equiv\OP^2/\muu^2$ and $\rrr\equiv\omega\G/\muu^2$.
The strong damping condition reads simply $\rrr\gg\sin2\mix$. In the interesting
region $\chi\lesssim 6\times 10^{-6}$ this is typically fulfilled  because
$\rrr \simeq \rr \simeq 10^{-5}(\omega/{\rm keV})$. At the low energy end of the spectrum we will
typically violate the strong damping
regime, but this region contributes comparatively little to the energy density.

The physical meaning of \eqref{strongD} is clear regarding \eqref{sto}. Since $D$ damps
$\mathbf{P}_\mathrm{T}$,
$\mathbf{P}$ will rapidly become attached to the $z$ axis (regardless of the initial value).
Note that, for $|\mathbf{V}_\mathrm{T}|\equiv\Delta\sin2\mixe=0$, $\mathbf{P}_\mathrm{T}=0$ is a
\emph{stationary} solution of \eqref{sto} for which the production rate would vanish.
Now let us switch on a small $\mathbf{V}_\mathrm{T} \neq 0$. The precession caused by
$\mathbf{V}_\mathrm{T}$ tries to move $\mathbf{P}$ away from the $z$-axis. However, any transverse
component is immediately damped away again because $D\gg|\mathbf{V}_\mathrm{T}|$.
While the precession keeps $|\mathbf{P}|$ constant, it is decreased by the damping.
Therefore, while the direction of $\mathbf{P}$ will remain roughly constant, its modulus will
slowly decrease in time and we approach `flavor equilbrium' at a rate
\eqref{strongD}.

Integrating Eqs. \eqref{sto}, \eqref{strongD} over time we can calculate the $\gamma\leftrightarrow\gp$
transition probability,
\begin{equation}
P_{\gamma\to\gamma^{\prime}}(T_f)=\frac{1}{2}\left(1-\mathrm{exp}\left\{ {\int_{\infty}^{T_f}
\frac{\overline\Gamma}{H}\frac{dT}{T}}\right\}\right)\ , \label{PT}
\end{equation}
where we have used a parametrisation in terms of the temperature $T$ instead of time.
Since the relevant part of the integral comes from the vicinity of the resonance, we expand the denominator
of $\overline\Gamma$ around $y=1$ up to the relevant quadratic term. Expressed in terms of $z=\sqrt[3]{y}$,
and using finally $\sin2\mix\simeq 2\mix$, the integral is
\begin{eqnarray}
&&\!\!\!\!\!\!\!\!\!\!\!\!\frac{1}{2}\int \frac{\overline \Gamma}{H T} dT \simeq
\mix^2\left[\frac{\G}{H T}\right]^{^{\R}} \hspace{-3mm}\int \frac{T^\R dz}{\rr^2+9(z-1)^2} \\\nonumber
&&\simeq \mix^2\left[\frac{\G}{H}\right]^\R\hspace{-2mm}\frac{\pi}{3 \rr} =
\frac{\pi}{3}\frac{\muu^2\mix^2}{H^\R \omega} = \frac{1.0 \times 10^{11}\ \mix^2}{w} \ ,
\end{eqnarray}
where we have evaluated the braced combination at $T^\R$ (since it does not depend on $T$) and used
$w=\omega/T^\R$.

Our main result, namely  the fraction of the energy stored in the hCMB, is obtained by integrating
the probability
over the Boltzmann distribution,
%
\begin{eqnarray}
x=\frac{15}{\pi^4}\int_{w_0}^\infty\frac{w^3}{e^{w}-1}P_{\gamma\to\gamma^{\prime}}.
\end{eqnarray}
To be conservative we shall cut off the integral at small frequencies where the strong damping approximation
breaks down,
$\rr \simeq 2\mix$, corresponding to $\omega_0\simeq 2\times 10^{5}\mix$ keV.
If $\mix$ is large, this can affect a sizable part of the spectrum. Otherwise, we obtain an
analytic expression by expanding \eqref{PT},
\begin{eqnarray}
\label{estimate}
x \simeq 3.9 \times 10^{10}\, \mix^2 \ \ .
\end{eqnarray}

During this epoch, double Compton scattering is active,
since $\Gamma_\mathrm{2C}\sim \alpha\G(\omega/m_e)^2$ is
typically $\gg H^{\R}$ (for $m_{\gamma^{\prime}}\gtrsim 0.1$ meV). Therefore the remaining photons will
regain a black body spectrum after the resonant production.\\

\noindent \emph{Discussion and conclusions.-}
In the region $0.1\,{\rm meV}\lesssim \muu\lesssim 10\,{\rm meV}$, the fraction of photons converted
into hidden photons depends only very weakly on the mass (cf. Eq.~\eqref{estimate} and Fig.~\ref{Prob}).
For $\mix\gtrsim 10^{-5}$, the resonant production would be sufficiently strong to convert roughly half the
photon energy into hidden photons. This would clearly be in contradiction with cosmological observations:
even the conversion of a small part of the photon energy into hidden photons may leave
observable traces in the effective number of relativistic species $N_\nu^\mathrm{eff}$ and
in the baryon to photon ratio $\eta$.
Using $x<0.2$, from the upper limit on $N_\nu^\mathrm{eff}$  in Eq.~\eqref{Neff}, or the slightly
weaker constraint $x\lesssim 0.32$, from the upper limit  on $\eta$ in Eq.~\eqref{etaBBN},
one obtains an upper bound $\chi\lesssim (3-4)\times 10^{-6}$, in the mass range
$0.1\,{\rm meV}\lesssim \muu\lesssim 10\,{\rm meV}$ (cf.  Fig.~\ref{Prob}).
In the mass range
$0.15\,{\rm meV}\lesssim m_{\gamma^{\prime}}\lesssim 0.3 \,{\rm meV}$,
this improves upon the previously established upper bounds on $\chi$
-- from searches for deviations from the Coulomb law and from
light-shining-through-walls (LSW) experiments -- by a factor of up to $2$ (cf.  Fig.~\ref{Prob}).

As mentioned earlier, inclusion of Lyman-$\alpha$ data seems to favor higher values of
$N_\nu^\mathrm{eff}>3$ \cite{Seljak:2006bg,Hamann:2007pi}.
It is tempting to interpret this in terms of an hCMB
(for an alternative explanation see \cite{Ichikawa:2007jv}).
Fortunately, the parameter region corresponding to the required energy fraction, $x\simeq 0.1$,
can be explored in the near future by pure laboratory experiment, notably
by LSW experiments \cite{Ahlers:2007qf} or an experiment exploiting microwave
cavities \cite{Jaeckel:2007ch} (cf.  Fig.~\ref{Prob}).
Furthermore,
the hidden photon CMB itself could be tested by an experiment like ADMX \cite{Asztalos:2001tf}
in which hidden photons entering a cavity
can be reconverted into detectable ordinary photons.

\vspace{-3ex}

\begin{thebibliography}{99}

\vspace{-3ex}
\bibitem{Holdom:1985ag}
  B.~Holdom,
  Phys.\ Lett.\  B {\bf 166}, 196 (1986);
  R.~Foot, and X.~G.~He,
  Phys.\ Lett.\  B {\bf 267}, 509 (1991).

\bibitem{Dienes:1996zr}
  K.~R.~Dienes {\em et al.},
  Nucl.\ Phys.\  B {\bf 492}, 104 (1997);
  S.~A.~Abel and B.~W.~Schofield,
  Nucl.\ Phys.\  B {\bf 685}, 150 (2004);
  S.~A.~Abel {\em et al.},
  hep-ph/0608248;
  0803.1449 [hep-ph].

\bibitem{Okun:1982xi}
  L.~B.~Okun,
  Sov.\ Phys.\ JETP {\bf 56}, 502 (1982).


\bibitem{Davidson:2000hf}
  S.~Davidson, S.~Hannestad and G.~Raffelt,
  JHEP {\bf 0005}, 003 (2000).

\bibitem{Ichikawa:2008pz}
  K.~Ichikawa {\em et al.},
  0803.0889 [astro-ph];
  J.~Dunkley {\it et al.}  [WMAP],
  0803.0586 [astro-ph].

\bibitem{Mangano:2006ur}
  G.~Mangano {\it et al.},
  JCAP {\bf 0703}, 006 (2007);
  K.~Ichikawa, M.~Kawasaki and F.~Takahashi,
  JCAP {\bf 0705}, 007 (2007);
  E.~Komatsu {\it et al.}  [WMAP],
  0803.0547 [astro-ph].

\bibitem{Seljak:2006bg}
  U.~Seljak {\em et al.},
  JCAP {\bf 0610}, 014 (2006);
  M.~Cirelli and A.~Strumia,
  JCAP {\bf 0612}, 013 (2006).

\bibitem{Hamann:2007pi}
  J.~Hamann {\em et al.},
  JCAP {\bf 0708}, 021 (2007).

\bibitem{Simha:2008zj}
  V.~Simha and G.~Steigman,
  0803.3465 [astro-ph].

\bibitem{Harris:1980zi}
  R.~A.~Harris and L.~Stodolsky,
  Phys.\ Lett.\  B {\bf 116}, 464 (1982).

\bibitem{Stodolsky:1986dx}
  L.~Stodolsky,
  Phys.\ Rev.\  D {\bf 36}, 2273  (1987).

\bibitem{Raffelt:1996wa}
  G.~G.~Raffelt,
  ``Stars As Laboratories For Fundamental Physics''
{\it  Chicago, USA: Univ. Pr. (1996) 664 p}

\bibitem{Bartlett:1988yy}
  D.~F.~Bartlett and S.~Loegl,
  Phys.\ Rev.\ Lett.\  {\bf 61}, 2285 (1988);
  E.~R.~Williams, J.~E.~Faller and H.~A.~Hill,
  Phys.\ Rev.\ Lett.\  {\bf 26}, 721 (1971).

\bibitem{Redondo:2008aa}
  J.~Redondo,
  0801.1527 [hep-ph].

\bibitem{Ahlers:2007qf}
  M.~Ahlers {\it et al.},
  0711.4991 [hep-ph], and refs. therein.

\bibitem{ALPS}
 See http://alps.desy.de/e55 .

\bibitem{Jaeckel:2007ch}
  J.~Jaeckel and A.~Ringwald,
  Phys.\ Lett.\  B {\bf 659}, 509 (2008).

\bibitem{Gninenko}
  S.~Gninenko and J.~Redondo, 0804.3736 [hep-ex].

\bibitem{Ichikawa:2007jv}
  K.~Ichikawa {\it et al.},
  JCAP {\bf 0705}, 008  (2007).

\bibitem{Asztalos:2001tf}
  S.~Asztalos {\it et al.},
  Phys.\ Rev.\  D {\bf 64}, 092003 (2001).
\end{thebibliography}
\end{document}